\documentclass{article}

\usepackage{PRIMEarxiv}

\usepackage[utf8]{inputenc} 
\usepackage[T1]{fontenc}    
\usepackage{hyperref}       
\usepackage{url}            
\usepackage{booktabs}       
\usepackage{amsfonts}       
\usepackage{nicefrac}       
\usepackage{microtype}      
\usepackage{lipsum}
\usepackage{fancyhdr}       
\usepackage{graphicx}       
\graphicspath{{media/}}     

\title{IMCDCF: An Incremental Malware Detection Approach Using Hidden Markov Models}

\author{
  Ran Liu \\
  Univ. of Maryland, Baltimore County \\
  \texttt{\{rliu2@umbc.edu} \\
   \And
  Charles Nicholas \\
  Univ. of Maryland, Baltimore County \\
  \texttt{nicholas@umbc.edu} \\
}

\begin{document}
\maketitle

\begin{abstract}
Dynamic malware analysis has become popular because it allows analysts to observe the behavior of running samples, facilitating improved decisions for malware detection and classification.  With the increasing number of new malware, there is a growing need for an automated malware analysis engine that can accurately detect malware samples.  In this paper, we briefly introduce the malware detection and classification approaches. Furthermore, we introduce a new malware detection and classification framework that works specifically in the dynamic analysis setting, namely Incremental Malware Detection and Classification Framework, or IMDCF. In this paper, we present a novel framework designed specifically for the dynamic analysis setting, named the Incremental Malware Detection and Classification Framework (IMDCF). IMDCF provides a end-to-end solution for general-purpose malware detection and classification with 96.49\% accuracy and simple architecture.
\end{abstract}

\section{Introduction}
Malware has long been a prevalent security threat globally. There have been many approaches for malware detection that can be categorized into the following groups: static malware analysis, dynamic malware analysis, and machine learning malware analysis\cite{Shalaginov2018MachineTutorial}\cite{Nath2014StaticMethods}. Static malware analysis involves examining the executable binary or source code. However, this approach has its limitations, as source code may not always be accessible. In contrast, dynamic malware analysis investigates the malware as it runs, often in a sandbox environment, and is also known as behavior-based malware analysis\cite{Egele2012ATools}\cite{Or-Meir2020DynamicSurvey}. In our work, we employ Hidden Markov Models (HMMs) to clarfify malware families, including a general "benign" family for benign specimens. In the IMDCF framework, each malware and benign family is used to train a set of corresponding HMMs. Each HMM within the same family is trained on a disjoint subset of the family dataset, covering at most 30\% of the total features. The test sample is evaluated independently by HMMs from the same family, and the mean of their result scores is assigned to the sample, indicating the likelihood that the HMMs trained on that family would accept it. An input specimen is considered to belong to a specific malware or benign family if the computed probability is high. However, since high probability data items may be produced by different malware or benign families, IMDCF forms longer data sequences by combining previously seen input data items. The newly-formed longer data sequence is scored against each HMM family, and the sequence is assigned to the malware or benign family with the highest score. If no model accepts the given data sequence, it is considered part of a new malware family, potentially requiring the training of a new HMM for that family. IMDCF has several advantages:
\begin{itemize}
    \item High accuracy: IMDCF achieves up to 96.49\% classification accuracy when presented with mixed malware types.
    \item Incremental processing: IMDCF works with sequences of data, handling items as short as one opcode or system call while maintaining high accuracy.
    \item Versatility: IMDCF is designed for use in various settings, including mobile and IoT environments, and can be applied for purposes such as anomaly detection.
    \item Simplicity: IMDCF features a relatively simple structure, allowing for easy implementation and adoption.
\end{itemize}

The paper is organized as follows: The related works are introduced in Section 2. The HMM's background is introduced in Section 3. In section 4, we describe and introduce our framework. In section 5, we'll discuss our experiment. The future work is discussed in Section 6. 

\section{Related Work}
Konrad Rieck et al. proposed an incremental malware analysis framework that incrementally extracts prototypes from test samples\cite{RieckAutomaticLearning}. Their framework starts by running and observing malware activity in a sandbox, generating a report containing running behavior. This report is then embedded into a higher-dimensional vector, with each dimension representing a similar behavior pattern. Machine learning techniques, such as KMM, are applied to embedded reports to cluster and classify malware samples incrementally, for instance, on a daily basis. New prototype classes are subsequently added for further analysis. In comparison to Rieck's work, IMDCF works on dynamic features that incrementally feeding the classification engine with running behaviors, enabling classification while the malware is still active in the sandbox. 

Shraddha Suratkar et al. investigated the use of HMM in anomaly detection\cite{8973098}. Machine learning techniques are initially applied to test samples for anomaly detection, with trained HMMs then utilized to predict the next most probable system calls during an attack. Jing Zhao et al. discussed the efficiency of using a Gaussian Mixture Hidden Markov Model for malware classification\cite{Zhao2021MalwareCW}. Iyer, Divya et al. employed HMM for credit card anomaly detection, using transaction categories as hidden states and transaction history as the observation sequence\cite{6141395}. By computing the probability of a new transaction being accepted by a given HMM, they determine whether the transaction is fraudulent. IMDCF employs a similar approach to compute the probability of new features.

\section{Hidden Markov Model}
HMM (Hidden Markov Model)\cite{Juang1984OnView} is a statistical model and is especially useful for modeling time series data or sequences where the underlying process generating the observations is supposed to be a Markov process with unobservable (hidden) states, which can be characterized as follows\cite{Rabiner1989}:
\begin{itemize}
    \item Hidden States of Markov process $S = \{s_1,s_2,...,s_n\}$, and $q_t$ donates the state at time instant t.
    \item A set of possible observations $V = \{V_1,V_2,...,V_m\}$, where m is the number of distinct symbols for each state.
    \item The state transition probabilities stochastic matrix $A = [a_{i,j}]$. $a_{ij} = P(q_{t + 1} = s_j|q_{t} = s_i)$, where $1 \leq i \leq N$, $1 \leq j \leq N$ N is the number of states of the model.
    \item The observation transition probabilities stochastic matrix $B = [b_j(k)]$. $b_j(k) = P[V_k|s_j]$,  where $1 \leq i \leq N$, $1 \leq j \leq M$.
    \item The initial state distribution vector $[\pi_i]$. $\pi_i = P[q_1 = s_i]$,where $1 \leq i \leq N$.
    \item The observation sequence $O = \{O_1, O_2,...,O_R\}$, where R is length of the observation sequence.
\end{itemize}

Assuming we have a HMM $\lambda(A,B,\pi)$ of a malware family, to decide if the fed observation sample  $O = O_1,O_2,...,O_t$ is belonging to the malware family can be computed as follows:
\begin{equation}
    P(O|\lambda) = \sum_{Q}P(O|Q,\lambda)P(Q|\lambda)
\end{equation}
, where $Q = q_1,q_2,...,q_R$ is the optimal states, $P(Q|\lambda)$ is the probability of Q by given HMM $\lambda(A,B,\pi)$ and $P(O|Q,\lambda)$ is the probability of observations $O = O_1,O_2,...,O_t$ is generated by states $Q = q_1,q_2,...,q_R$. 

We have:
\begin{equation}
   P(O|Q,\lambda) = \prod_{t=1}^R P(O_t|q_t,\lambda) = b_{q_1}(O_1)b_{q_2}(O_2)...b_{q_R}(O_R).
\end{equation}

We further know:
\begin{equation}
    P(Q|\lambda) = \pi_{q_1}a_{q_1q_2}a_{q_2q_3}...a_{q_{R-1}q_{R}}
\end{equation}

We say observations $O$ is accepted by the HMM $\lambda$ with low probability if $P(O|\lambda) \leq Threshold $
As introduced in \cite{6141395}: Assume we have an observation sequence text$O = O_1,O_2,...,O_t$ at time t that $lambda$ has accepted, and a new observation $O_{t+1}$ at time t+1. We remove $O_1$ from $O$ and add $O_{t+1}$ to $O$ to form a new sequence: $Oprime = O_2,O_3,...,O_{t+1}$. $O_{t+1}$ has a low likelihood of being accepted given HMM if $P(Oprime|lambda) - P(O|lambda) leq 0$. then this sample belongs to a new malware family. Otherwise, we construct a longer sequence $Oprime = O_{t+1},O_{t+2},...,O_{t+2000}$ and compute $Oprime$ scores of all HMMs. The final decision is made by a majority vote.

\section{Incremental Malware Classification Framework Description} \label{sec:results}
\begin{figure*}[!t]
    \centering
    \includegraphics[width = 5in]{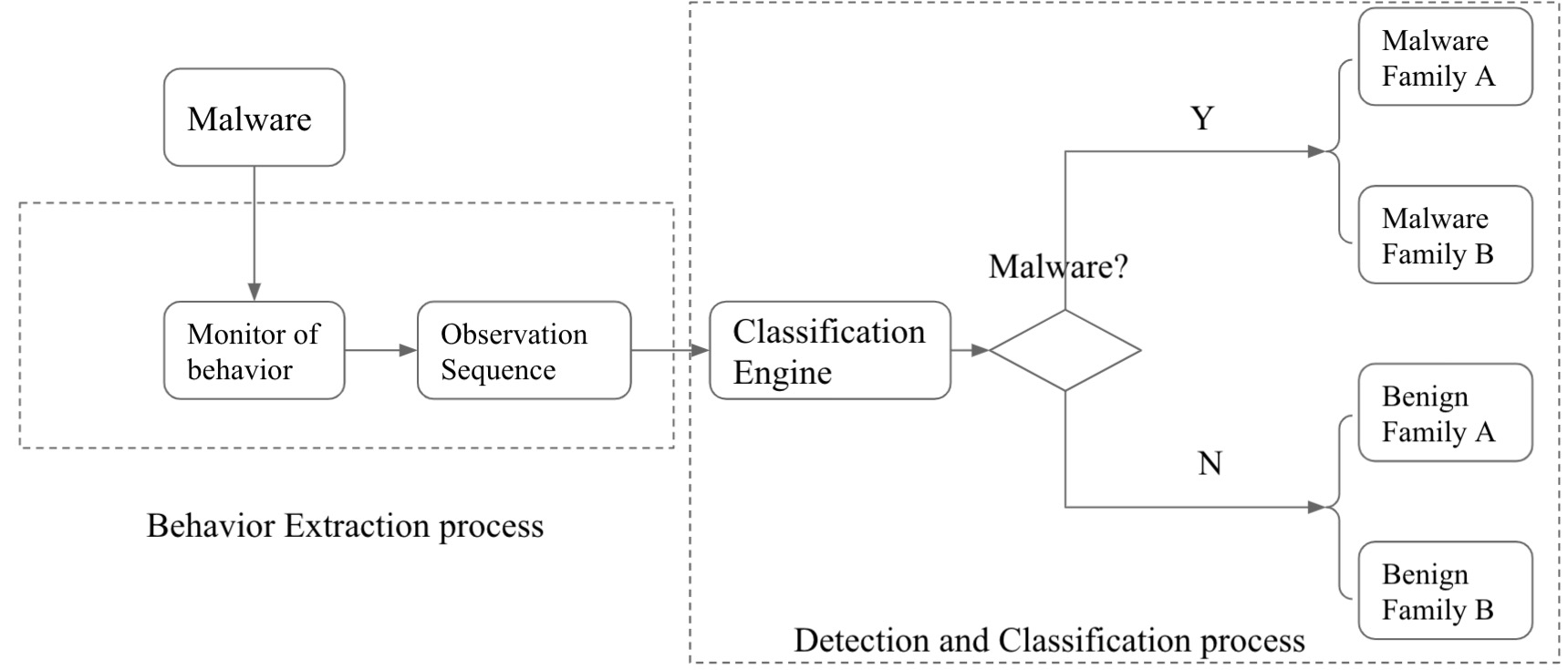}
    \caption{Incremental Malware Classification Framework contains two different process. The Malware first is monitored and executed in the sandbox, which calls behavior extraction process. The classification process has trained HMMs. The training process is done offline. Each HMM evaluates the coming observation sequence and assigns score to it.}
    \label{fig:framework}
\end{figure*}

IMDCF contains two processes: the Behavior Extraction process and Detection process. A schematic overview of IMDCF is shown in \autoref{fig:framework}

IMDCF begins by gathering running behavior in the sandbox, such as opcodes and API calls. The opcodes are then sorted by frequency and encoded into 26 alphabetic symbols, with all other opcodes encoded as the special symbol '$\bigcap$'. For example, sequence $MOV -> PUSH -> ADD -> SUB$ is encoded to $A -> B -> C -> D$.

A group of one class HMMs is trained, and each is trained with either benign files or one of the malware families. At time t, the sample generates an opcode $O_t$. For distinct malware families and benign families, an initial sequence of opcodes is prepared. For instance, family A has a sequence $O_a = O_1, O_2, ..., O_{t-1}$, while family B has a sequence $O_b = O_1', O_2', ..., O_{t-1}'$. For each sequence, a new observation sequence is constructed by dropping $O_1$ (or $O_1'$) and appending $O_t$ to the respective sequences. This process updates the observation sequences for each family, reflecting the most recent opcode information. Each HMM then measure the likelihood of accept the new generating sequence.
\begin{itemize}
\item If the log likelihood of sequence $O$ is below the threshold for all HMMs, the sample generating $O_t$ is considered to belong to a new malware family.
\item If the log likelihood one sequence $O$ exceeds the threshold of an HMM, the sample generating $O_t$ is considered to belong to that malware family.
\item If the log likelihood one sequence $O$ exceeds the threshold of more than one HMM, a longer sequence $O = O_t O_{t+1}...$ is constructed. Repeating above process until only one HMM left.
\end{itemize}
\section{Experiment}
In our experiments, We analyze two malware families, Zeroaccess and Zbot and a benign family.
We collect a set of opcodes from benign software without grouping them into families, as our primary concern is whether IMDCF can detect malware among the samples. The Zeroaccess is a Windows System Malware primarily used for downloading other malware samples onto infected computers. We collect 1,308 Zeroaccess files. Zbot is a Windows System Malware primarily used for stealing banking information. and we collect 2,136 Zbot files. Each HMM is trained with a subset of 20 files from the respective malware family, with observation sequences of at least around 100,000. To achieve the best training results, we set the iteration count to 200.
\subsection{Malware Detection Experiment}
\begin{figure}[!t]
    \centering
    \includegraphics[width = 3.0in]{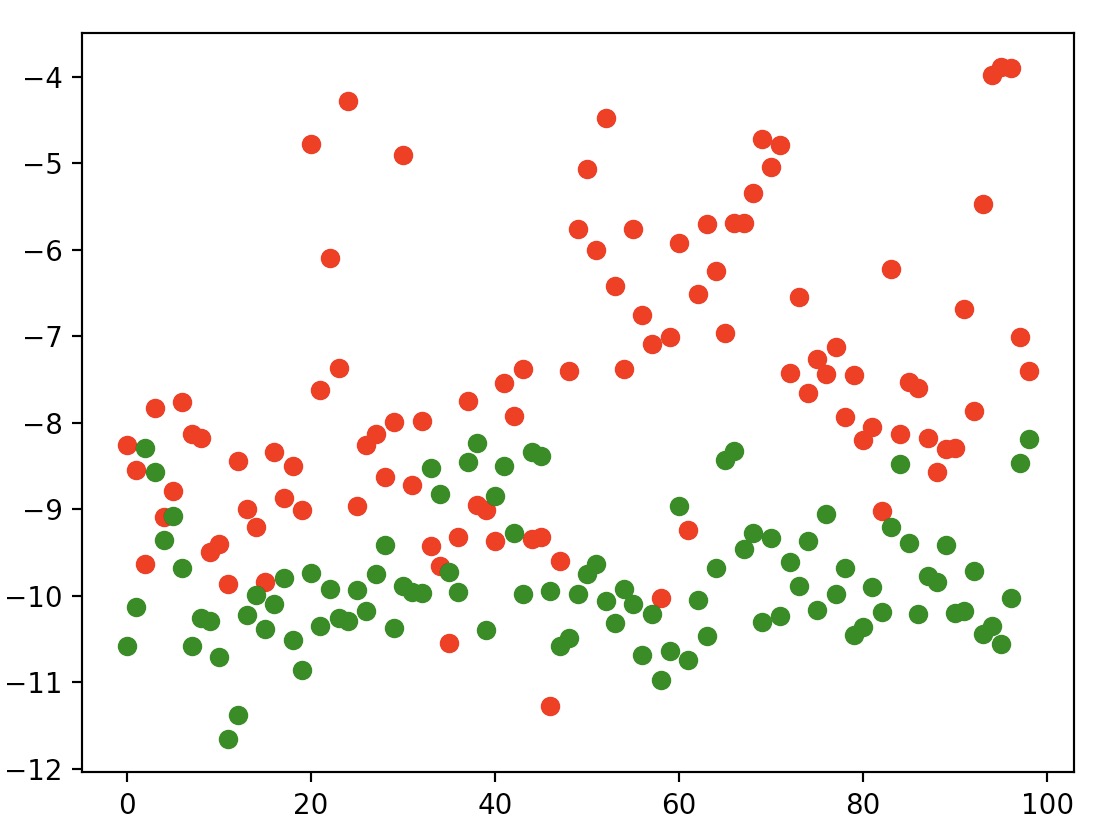}
    \caption{Plot shows the IMDCF detection results with Log likelihood per opcode (LLPO) as y-axis and sample index as x-axis. }
    \label{fig:benign}
\end{figure}

IMDCF can achieve 0.9091 accuracy score. As shown in \autoref{fig:benign}, IMDCF can successfully detect malware while producing few false positives

\begin{figure}[!ht]
    \centering
    \includegraphics[width = 3.0in]{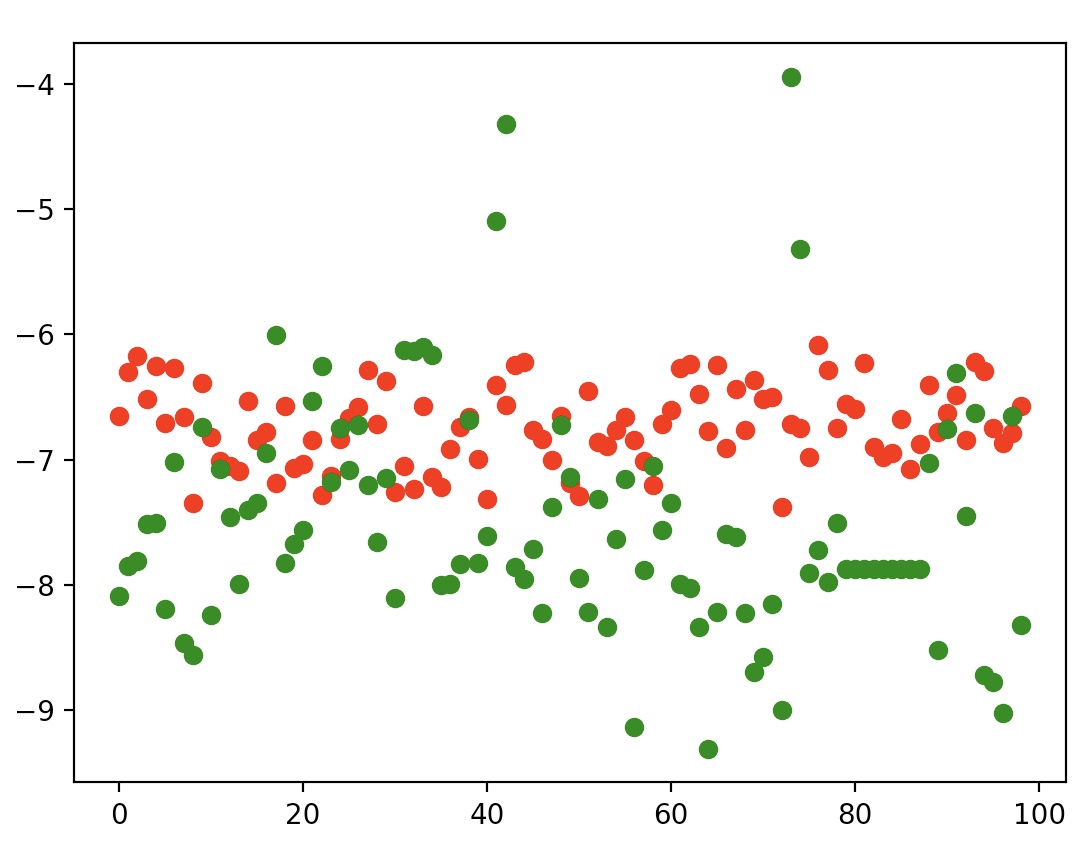}
    \caption{Plot shows the IMDCF classification results using Zeroaccess Model with Log likelihood per opcode (LLPO) as y-axis and sample index as x-axis. }
    \label{fig:zeroaccess}
\end{figure}

\begin{figure}[!ht]
    \centering
    \includegraphics[width = 3.0in]{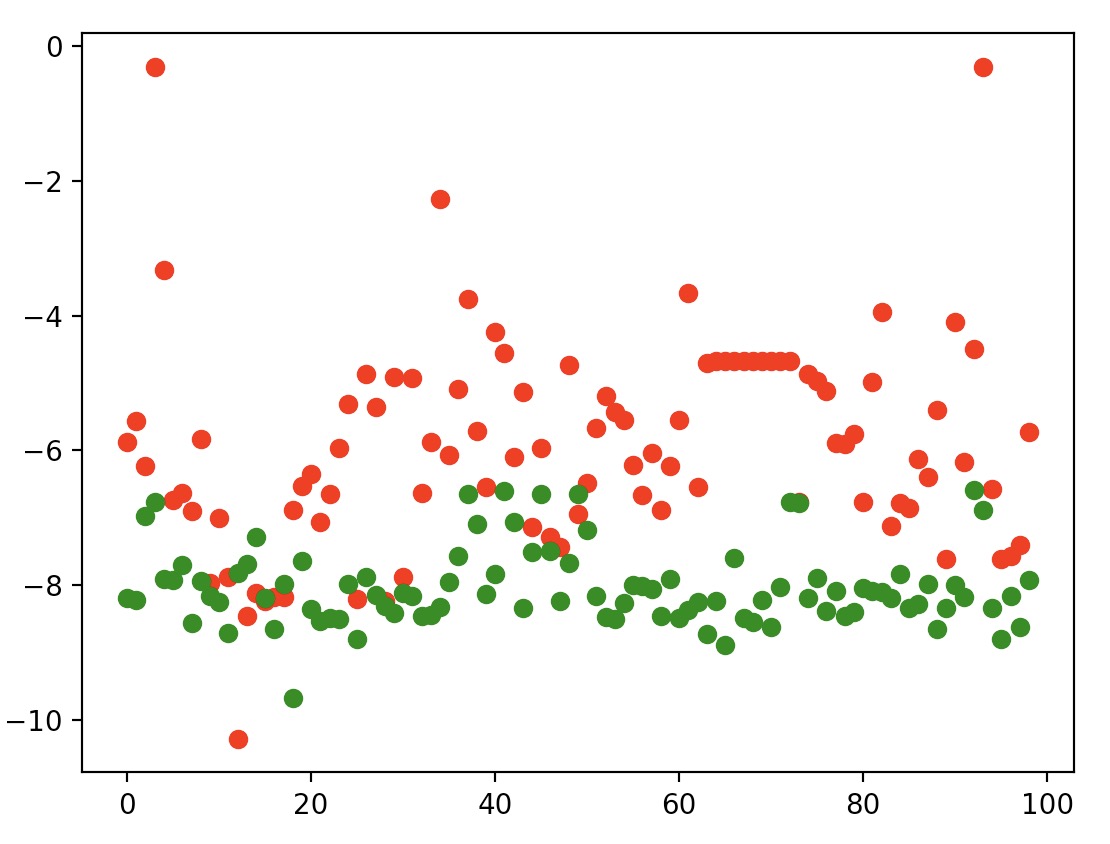}
    \caption{Plot shows the IMDCF classification results using Zbot Model with Log likelihood per opcode (LLPO) as y-axis and sample index as x-axis. }
    \label{fig:zbot}
\end{figure}

The \autoref{fig:zeroaccess} and \autoref{fig:zbot} show the classfication results using Zeroaccess and Zbot model with the accuracy score 0.8384.


\begin{figure}
    \centering
    \includegraphics[width = 3.0in]{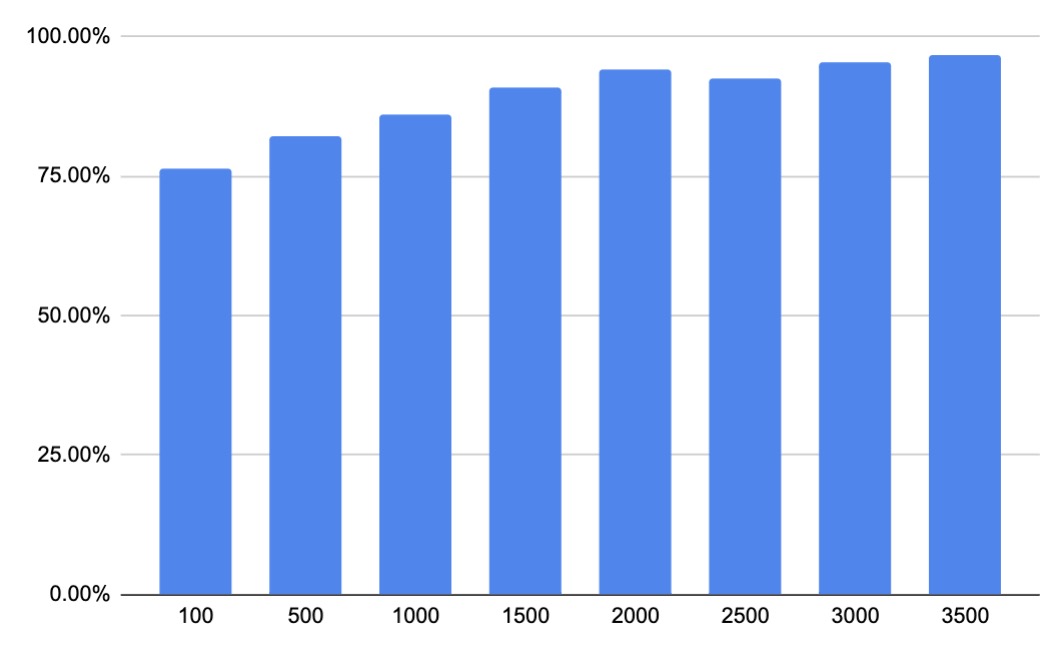}
    \caption{Plot shows accuracy score with increasing observation length. }
    \label{fig:run}
\end{figure}

The observation sequence length can affect the classification accuracy. As shown in \autoref{fig:run}, Th accuracy increases as the length increases.

\section{Conclusion}
Malware addresses a lot of attention from both industry and academic since it’s a major threat for today's network security. In this work, we present the IMDCF, the incremental malware classification model under dynamic analysis setting. For the malware detection task, IMDCF can use only one opcode with comparable accuracy. We are excited about the possibility of detecting malware using short sequences and plan to apply this model to other tasks.

\bibliographystyle{unsrt}
\bibliography{references, Mendeley}

\begin{thebibliography}{10}

\bibitem{Shalaginov2018MachineTutorial}
Andrii Shalaginov, Sergii Banin, Ali Dehghantanha, and Katrin Franke.
\newblock {Machine Learning Aided Static Malware Analysis: A Survey and
  Tutorial}.
\newblock pages 7--45. 2018.

\bibitem{Nath2014StaticMethods}
Hiran~V. Nath and Babu~M. Mehtre.
\newblock {Static Malware Analysis Using Machine Learning Methods}.
\newblock pages 440--450. 2014.

\bibitem{Egele2012ATools}
Manuel Egele, Theodoor Scholte, Engin Kirda, and Christopher Kruegel.
\newblock {A survey on automated dynamic malware-analysis techniques and
  tools}.
\newblock {\em ACM Computing Surveys}, 44(2):1--42, 2 2012.

\bibitem{Or-Meir2020DynamicSurvey}
Ori Or-Meir, Nir Nissim, Yuval Elovici, and Lior Rokach.
\newblock {Dynamic Malware Analysis in the Modern Era—A State of the Art
  Survey}.
\newblock {\em ACM Computing Surveys}, 52(5):1--48, 9 2020.

\bibitem{RieckAutomaticLearning}
Konrad Rieck, Philipp Trinius, Carsten Willems, and Thorsten Holz.
\newblock {Automatic Analysis of Malware Behavior using Machine Learning}.
\newblock Technical report.

\bibitem{8973098}
Shraddha Suratkar, Faruk Kazi, Rohan Gaikwad, Akshay Shete, Raj Kabra, and
  Shantanu Khirsagar.
\newblock Multi hidden markov models for improved anomaly detection using
  system call analysis.
\newblock In {\em 2019 IEEE Bombay Section Signature Conference (IBSSC)}, pages
  1--6, 2019.

\bibitem{Zhao2021MalwareCW}
Jing Zhao, Samanvitha Basole, and M.~Stamp.
\newblock Malware classification with gmm-hmm models.
\newblock {\em ArXiv}, abs/2103.02753, 2021.

\bibitem{6141395}
Divya. Iyer, Arti Mohanpurkar, Sneha Janardhan, Dhanashree Rathod, and Amruta
  Sardeshmukh.
\newblock Credit card fraud detection using hidden markov model.
\newblock In {\em 2011 World Congress on Information and Communication
  Technologies}, pages 1062--1066, 2011.

\bibitem{Juang1984OnView}
B.-H. Juang.
\newblock {On the Hidden Markov Model and Dynamic Time Warping for Speech
  Recognition-A Unified View}.
\newblock {\em AT{\&}T Bell Laboratories Technical Journal}, 63(7):1213--1243,
  9 1984.

\bibitem{Rabiner1989}
L.R. Rabiner.
\newblock {A tutorial on hidden Markov models and selected applications in
  speech recognition}.
\newblock In {\em Proceedings of the IEEE ( Volume: 77, Issue: 2, Feb. 1989)},
  pages 257 -- 286. IEEE, 2 1989.

\end{thebibliography}

\end{document}